# Thermal hysteretic behavior and negative magnetoresistance in an unusual charge-density-wave material EuTe$_4$

Q. Q. Zhang[1,2*], Y. Shi[3*], K. Y. Zhai[1,2], W. X. Zhao[1,2], X. Du[1,2], J. S. Zhou[1,2], X. Gu[1,2], R. Z. Xu[1,2], Y. D. Li[1,2], Y. F. Guo[4,5], Z. K. Liu[4,5], C. Chen[4,5,6], S.-K. Mo[6], T. K. Kim[7], C. Cacho[7], J. W. Yu[1,2], W. Li[1,2], Y. L. Chen[4,5,8], Jiun-Haw Chu[3*], and L. X. Yang[1,2*]

[1]*State Key Laboratory of Low Dimensional Quantum Physics, Department of Physics, Tsinghua University, Beijing 100084, China.*
[2]*Frontier Science Center for Quantum Information, Beijing 100084, China.*
[3]*Department of Physics, University of Washington, Seattle, WA, 98195, USA*
[4]*School of Physical Science and Technology, ShanghaiTech University and CAS-Shanghai Science Research Center, Shanghai 201210, China.*
[5]*ShanghaiTech Laboratory for Topological Physics, Shanghai 200031, China.*
[6]*Advanced Light Source, Lawrence Berkeley National Laboratory, Berkeley, CA 94720, USA*
[7]*Diamond Light Source, Didcot OX11 0DE, UK.*
[8]*Department of Physics, Clarendon Laboratory, University of Oxford, Parks Road, Oxford OX1 3PU, UK.*

*Emails: JHC: jhchu@uw.edu; LXY: lxyang@tsinghua.edu.cn*

**EuTe$_4$ is a newly-discovered van der Waals material exhibiting a novel charge-density wave (CDW) with a large thermal hysteresis in the resistivity and CDW gap. In this work, we systematically study the electronic structure and transport properties of EuTe$_4$ using high-resolution angle-resolved photoemission spectroscopy (ARPES), magnetoresistance measurements, and scanning tunneling microscopy (STM). We observe a CDW gap of about 200 meV at low temperatures that persists up to 400 K, suggesting that the CDW transition occurs at a much higher temperature. We observe a large thermal hysteretic behavior of the ARPES intensity near the Fermi level, consistent with the resistivity measurement. The hysteresis in the resistivity measurement does not change under a magnetic field up to 7 T, excluding the thermal magnetic hysteresis mechanism. Instead, the surface topography measured with STM shows surface domains with different CDW trimerization directions, which may be important for the thermal hysteretic behavior of EuTe$_4$. Interestingly, we observe a large negative magnetoresistance at low temperatures that can be associated with the canting of magnetically ordered Eu spins. Our work shed light on the understanding of magnetic, transport, and electronic properties of EuTe$_4$.**

## I. INTRODUCTION

In solid materials, the interaction and cooperation between electrons and lattice can induce a long-range ordered ground state in which charge and lattice pattern in a new periodic structure that is different from the normal-state lattice structure [1,2]. Such a charge-density wave (CDW) state, since its discovery, has been extensively studied for decades [3-6]. While the consequence of symmetry breaking is evident, the mechanism of the CDW transition in each materials system varies and is often under debate. In a half-filled one-dimensional (1D) electronic system, the nuclear atoms are susceptible to the dimerization due to the perfectly nested Fermi surface (FS), which is accompanied by the fully opened energy gap in the electronic density of states and the softening of the phonons with the momentum of twice the Fermi momentum of the electronic energy band [1]. However, the FS nesting scenario has been challenged [7-10], particularly in many two-dimensional (2D) and 3D materials [11-14], where the FS is usually not perfectly nested, resulting in partial energy gaps and metallic CDW states [15-18].

The binary rare-earth polytellurides $R$Te$_n$ (where $R$ is a rare-earth element, n = 2, 2.5, 3, 4) provide an ideal platform to study CDW transition in van der Waals materials [15-24]. The common structure of these compounds consists of corrugated $R$-Te slabs and planar square Te sheets that alternately stack along the $c$-axis. The planar square Te nets are highly unstable towards lattice distortion, resulting in commensurate or incommensurate CDW states [25-27]. The 5$p$ orbitals of Te atoms in the planar Te nets are partially filled, constructing the electronic energy bands near the Fermi level ($E_F$). The $R$-derived states, on the other hand, usually locate below $E_F$ and make negligible contribution to the FS and CDW order. In general, the CDW transition in $R$Te$_n$ is interpreted by FS nesting, although most of them remain metallic in the

CDW state owing to the imperfect FS nesting [15-17,28].

Recently, a new type of rare-earth tetratelluride EuTe$_4$ was discovered to exhibit a non-metallic CDW state. It crystallizes into an orthorhombic lattice structure with space group of *Pmmn* (No.59) at room temperature. The lattice structure of EuTe$_4$ consists of corrugated Eu-Te layers separated by Te monolayers or bilayers, as schematically shown in Fig. 1(a) [21]. The quasi-square Te nets in EuTe$_4$ are unstable and prone to distort, leading to the trimerization of Te atoms. The CDW superstructure was initially determined to be $1a \times 3b \times 2c$ [21]. However, recent X-ray diffraction (XRD) experiment suggested an incommensurate CDW wave vector of $q_b = 0.643(3)b^*$, where $b^* = 2\pi/b$ [22]. The CDW transition temperature of EuTe$_4$ is still a mystery: while it was initially determined to be about 255 K from the resistivity measurement [21], recent angle-resolved photoemission spectroscopy (ARPES) and XRD measurements suggest that the CDW transition occurs at a much higher temperature above 400 K [22,23]. Density-functional theory (DFT) calculation suggests that the CDW transition is driven by FS nesting [29], despite that the crystal structure above 400 K remains unknown up to date. Interestingly, both the resistivity and ARPES measured CDW gap show a thermal hysteretic behavior in a large temperature range from 100 K to 400 K, which was attributed to the switching of the relative CDW phases in different Te layers [22]. However, compelling evidence of this phenomenological scenario is still lacking and the novel thermal hysteretic behavior is still not fully understood.

In this paper, we systematically study the electronic structure and electrical transport property of EuTe$_4$ using high-resolution ARPES, scanning tunneling microscopy (STM), and magnetoresistance measurement. XRD studies show no obvious change in the lattice structure in the temperature range from 125 K to 300 K except for a considerable thermal expansion. The ARPES intensity near $E_F$ shows a considerable thermal hysteresis, in good agreement with the resistivity measurement. The measurement of the resistivity under different magnetic fields excludes the thermal magnetic hysteretic behavior, while the sample surface topography unravels the coexistence of different domains of oppositely trimerized Te atoms in the same Te layer, which may play an important role in the thermal hysteretic behavior. Interestingly, we observe a negative magnetoresistance of about 86% under a magnetic field higher than 4 T at 2 K, which is consistent with the realignment of Eu ordered moments. Our results provide new insights into the understanding of novel properties of $EuTe_4$.

## II. METHOD

Single crystals of $EuTe_4$ were synthesized via a Te flux method [21]. Electrical transport measurements were carried out in a Quantum Design physical property measurement system (PPMS). Electrical contacts were prepared by attaching four Pt wires with silver-paste onto evaporated gold pads on the crystal surface. An 1 $mA$ current was applied in the ab-plane and the magnetic field was applied along the c-axis.

ARPES measurements were performed at beamline I05 of Diamond Light Source (DLS) [30], beamline 10.0.1 of the Advanced Light Source (ALS), beamline BL13U of National Synchrotron Radiation Laboratory (NSRL), and laser-based ARPES system in Tsinghua University. Data were collected by Scienta R4000 (DA30 L) analyzers at DLS and ALS (NSRL and Tsinghua University), respectively. Energy and angular resolutions were set to 15 meV (7 meV) and 0.2° for synchrotron (laser)-based experiments, respectively. The samples were cleaved *in situ* and measured under ultrahigh vacuum below $1 \times 10^{-10}$ Torr.

**III. RESULTS AND DISCUSSION**

Figure 1(b) shows the temperature-dependent resistivity measurement on our sample. The resistivity shows a non-metallic temperature dependence. We notice that low-temperature resistivity is considerably lower than the value in previous report [21] and can vary significantly depending on the growth conditions, which is typical for a nearly intrinsic semiconductor. Consistent with previous results, a hysteresis loop is observed in a large temperature range from about 100 K to 400 K [21]. Figures 1(c) and (d) show XRD patterns at 125 K and 300 K in the (*hk0*) plane, respectively. The superlattice spots exist at both 125 K and 300 K, implying a CDW state up to 300 K, consistent with previous results [22]. Therefore, the observed hysteretic behavior is not related to the CDW transition. We estimate a CDW wavevector of about 0.325 $b^*$ without noticeable change between 125 K and 300 K, which suggests an incommensurate CDW state. The observed shift of the scattering peaks suggests a large linear thermal expansion coefficient of about $2.24 \times 10^{-4}$ /K.

Figure 2 shows the basic electronic structure of EuTe$_4$. Figures 2(a) and (b) depict the 3D plot of ARPES spectra and constant energy contours (CECs) at selected binding energies, respectively. Due to the formation of CDW, the entire FS is gapped, inducing a semiconducting band structure that accords with the resistivity measurement [Fig. 1(b)]. The CECs shown in Fig. 2(b) exhibit zig-zag chains that can be well described by the tight-binding model [yellow and cyan curves in Fig. 2(b)], which is similar to the situation in *R*Te$_3$ [15-17] and suggests the relevance of FS nesting in the CDW transition [31]. With increasing binding energy, the CEC splits into parallel sheets. We observe weak states that can be connected to the main states by the CDW wave vector [red arrow in Fig. 2(b)], which are identified as CDW-folded states.

Figures 2(c)-(f) show the band dispersions along high symmetry directions collected with photon energy $hv$ = 130 eV at 12 K. Along $\bar{X}\bar{\Gamma}\bar{X}$ (cut #1), we observe linear-like band dispersions in a large energy range with band top at about -200 meV due to the formation of the CDW gap. By contrast, there exists no band dispersions along $\bar{Y}\bar{\Gamma}\bar{Y}$ (cut #2) down to -0.6 eV, suggesting a large anisotropy of the band structure. Along cut #3 and #4, we observe similar linear-like band dispersions with a CDW gap slightly smaller along $\bar{M}\bar{X}\bar{M}$. The CDW-folded bands are indicated by black arrows in Fig. 2(f), which show much weaker intensity compared with the main bands. Our experiment agrees with the previous results [22,23], despite the different photon energy used in our work. We note a band modulation near -0.5 eV as indicated by the red arrow in Fig. 2(c). This gap-like feature does not show a temperature dependence and exists only in the second Brillouin zone. It may be due to the existence of extra bands with band tops at about -0.7 eV and -0.55 eV (Supplemental Material, Fig. S1 [32]).

Figure 3 investigates the temperature evolution of the band structure using laser-ARPES. Figure 3(a) shows the electronic structure along the high symmetry $\bar{\Gamma}\bar{X}$ direction (cut #1) at selected temperatures. The band structure remains gapped up to 380 K, suggesting a much higher CDW transition temperature. Figure 3(b) shows the energy distribution curves (EDCs) integrated in a momentum window of about 0.08 Å$^{-1}$ around the momentum position of the valence band top, from which we observe the change of the peak position and width with temperature. Figure 3(c) shows the leading-edge gap (LEG) as a function of temperature. We observe a thermal hysteresis loop in the LEG, consistent with the previous ARPES measurement [22]. Nevertheless, the hysteresis loop is not so prominent as that in Ref. 22. The maximum difference between the LEG in the heating up and cooling down branches is about 10 meV, compared to a difference of about 40 meV in Ref. 22. We suggest that the quantitative difference from Ref. 22 may be due to the differences in samples. Indeed, our sample shows a lower resistivity at low temperatures and a less prominent hysteresis loop in the resistivity curve. Figure 3(d) shows the temperature evolution of ARPES intensity integrated in an energy window of 80 meV near $E_F$. We notice that ARPES intensity near $E_F$ shows a large thermal hysteresis loop with the intensity higher in the cooling branch, suggesting that the incoherent electronic states at $E_F$ is important for the resistivity and the thermal hysteretic behavior.

The thermal hysteretic phenomena are often observed in CDW materials featuring metastable states during the transition [33-37] and have been interpreted by different mechanisms such as the pinning of the CDW phase to impurities [34], metal-to-insulator transition [33], and incommensurate-to-commensurate CDW transition [37]. In EuTe$_4$, the hysteretic behavior was attributed to the switching of the relative phase of CDWs in different Te layers [22]. However, compelling evidence of such phenomenological scenario is still lacking.

Figure 4(a) schematically shows the crystal structures of $R$Te$_n$ compounds. They share the common structure units of $R$-Te slabs and nearly square Te nets that are energetically unstable to form the periodic distortion of Te atoms [21,25,26]. Among them, $R$Te$_2$ and $R$Te$_3$ feature only Te monolayers and bilayers, respectively, while EuTe$_4$ consists of both monolayers and bilayers of Te atoms. The coexistence of electrically neutral Te monolayers and bilayers in EuTe$_4$ was suggested to be crucial for the observed hysteretic behavior [22]. It is noticeable that no hysteretic behavior is observed in $R_2$Te$_5$ although it also features both Te monolayers and bilayers, which is possibly due to the fact that the Te layers in $R_2$Te$_5$ are not electrically neutral [38]. In addition, EuTe$_4$ also consists of unique Eu-Te slabs separated by Te monolayers, which may strongly affect the CDW structure and the hysteretic behavior.

On the other hand, the multi-$f$-electron nature and the magnetization of Eu atoms may also play a role in the hysteretic behavior. While the change of Eu valence has been ruled out by X-ray absorption near-edge spectroscopy experiments [22], the impact of the magnetization has yet not been investigated. In Fig. 4(b), we show the resistivity measurements under magnetic fields along the $c$-axis. We observe minor change of the hysteresis loop from 0 T to 7 T, which excludes the magnetization as the major mechanism of the hysteretic behavior [39,40].

Moreover, we also checked the CDW structure of the sample by using STM at 4 K. The STM topography in Fig. 4(c) visualizes the trimerization of Te atoms forming chain-like structures [21]. Noticeably, we observe CDW domains with opposite trimerizing directions as indicated by the green and orange rectangles in Fig. 4(c), suggesting the coexistence of out-of-phase CDWs in a single Te layer. Since the opposite trimerizations are energetically degenerate, they can coexist in both Te monolayers and bilayers, thus affecting the 3D stacking of the CDWs. We suggest that similar to the picture of magnetic domains and magnetic hysteresis loop, the competition of the CDW domains at different temperatures may play an important role in the thermal hysteretic behavior of EuTe$_4$.

Interestingly, in our magnetoresistance (MR) measurement, we observe a decrease of resistivity with increasing magnetic field as shown in Fig. 5(a). At 2 K, the in-plane resistivity decreases by about 86%, which is a considerable negative magnetoresistance (NMR) even larger than the value observed in Fe/Cr/Fe heterostructure (60%) [41]. The NMR effect is suppressed by increasing temperature and finally disappears at 25 K. Figures 5(b) and 5(c) show the magnetic susceptibility and magnetization measurement, respectively. The peak in the temperature dependence of magnetic susceptibility indicates that EuTe$_4$ orders antiferromagnetically below 7 K. From the magnetization versus field data, it can also be seen that the NMR effect is highly correlated with the canting of the Eu moment towards c-axis. Therefore, the NMR effect is very likely a result of the change of electronic structure due to the realignment of the spins of Eu$^{2+}$ ions. We notice that similar NMR effect has been observed in a related antiferromagnet EuTe$_2$, which was attributed to a similar spin-flip transition [42]. Magnetic ordering related large NMR was also observed in ferromagnetic Eu-based compounds EuIn$_2$P$_2$ and EuIn$_2$As$_2$ near their Curie temperatures [43,44].

## IV. CONCLUSION

In conclusion, we present a comprehensive investigation on the electronic structure and magnetoresistance of the van der Waals material EuTe$_4$. We observe a large thermal hysteresis in the resistivity and ARPES intensity near $E_F$. Several different mechanisms of the thermal hysteretic behavior are discussed. We suggest that the coexistence of opposite trimerization of Te atoms may play an important role in the thermal hysteretic behavior. Moreover, we reveal a large NMR effect in EuTe$_4$ at low temperatures, which may be due to the realignment of the spins of $Eu^{2+}$ ions. The semiconducting nature, thermal hysteretic behavior, and the large NMR effect make $EuTe_4$ an interesting platform for investigating unusual CDW transition in van der Waals materials.

## ACKNOWLEDGMENTS

This work was supported by the National Natural Science Foundation of China (Grant No. 12274251), the National Key R&D program of China (Grant No. 2017YFA0304600 and 2017YFA0305400), EPSRC Platform Grant (Grant No. EP/M020517/1). Y. F. G. acknowledges the support from the Natural Science Foundation of Shanghai (No. 17ZR1443300). L. X. Y. acknowledge the support from Tsinghua University Initiative Scientific Research Program. Materials synthesis and transport experiments at UW was supported by the David Lucile Packard Foundation and the Gordon and Betty Moore Foundation's EPiQS Initiative, Grant GBMF6759 to JHC. This research used resources of the Advanced Light Source, a US DOE Office of Science User Facility under Contract No. DE-AC02-05CH11231. We thank Diamond Light Source for access to beamline I05 (Proposal number SI22375-1) that contributed to the results presented here. We thank Prof. Zhe Sun for the support of experiments at beamline 13U of National Synchrotron Radiation Laboratory.

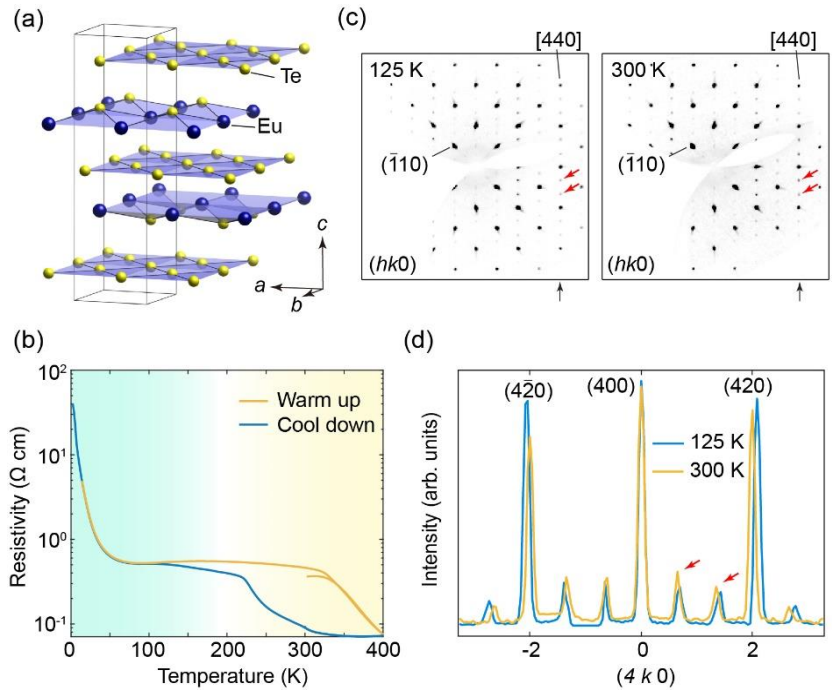

FIG.1. Basic characterization of EuTe$_4$ single crystal. (a) Illustration of the crystal structure of EuTe$_4$. (b) The temperature dependence of in-plane electrical resistivity. A thermal hysteresis loop is observed in a large temperature range. (c) X-ray diffraction (XRD) pattern along ($hk0$) at 125 K and 300 K, respectively. (d) Line profile of the XRD pattern along the [4$k$0] direction as indicated by the arrows in (c). The red arrows in (c) and (d) indicate the CDW superstructure spots and peaks.

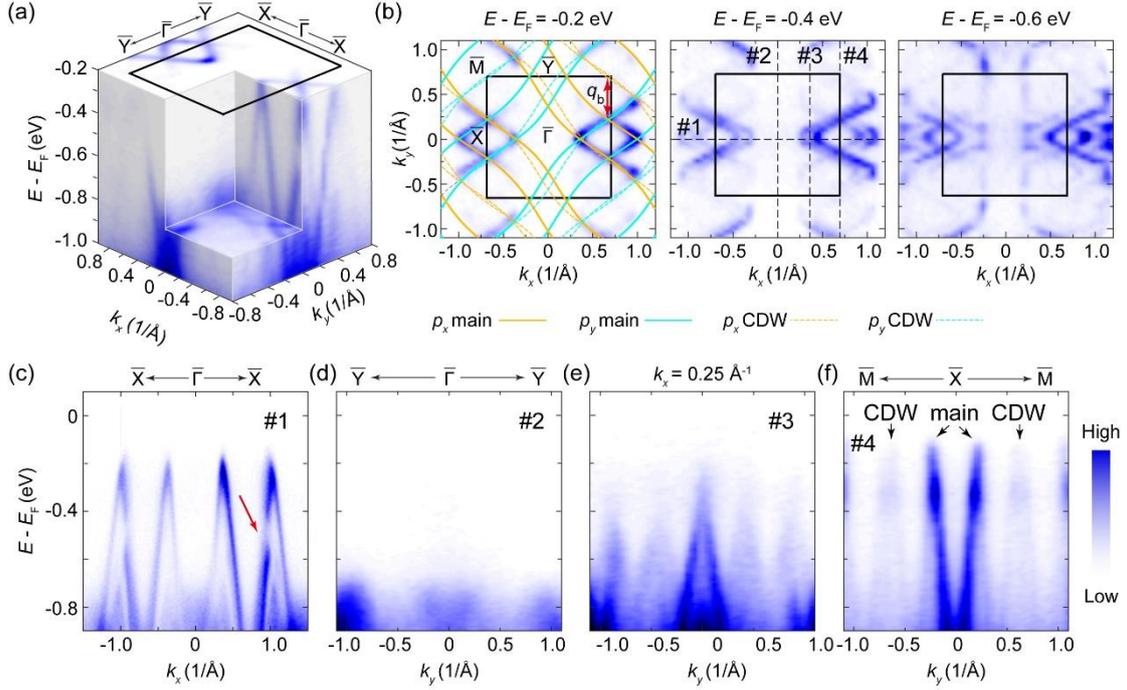

FIG.2. Basic electronic structure of EuTe$_4$. (a) Three-dimensional plot of the electronic structure of EuTe$_4$. The black lines indicate the surface Brillouin zone (BZ). (b) Constant energy contours (CECs) in the $k_x$-$k_y$ plane at selected binding energies, which were obtained by integrating ARPES intensity in an energy window of 100 meV. The solid black rectangle represents the surface BZ. The solid cyan and yellow lines are calculated band dispersions by tight-binding model, while the dashed cyan and yellow lines indicate the CDW-folded bands. (c-f) The band dispersion along cut #1-#4 as indicated by the dashed lines in (b). The red arrow in (c) indicates the band modulation near -0.5 eV. The main band and folded band are indicated by the black arrows in (f). Data were collected at 12 K with photon energy $hv$ = 130 eV.

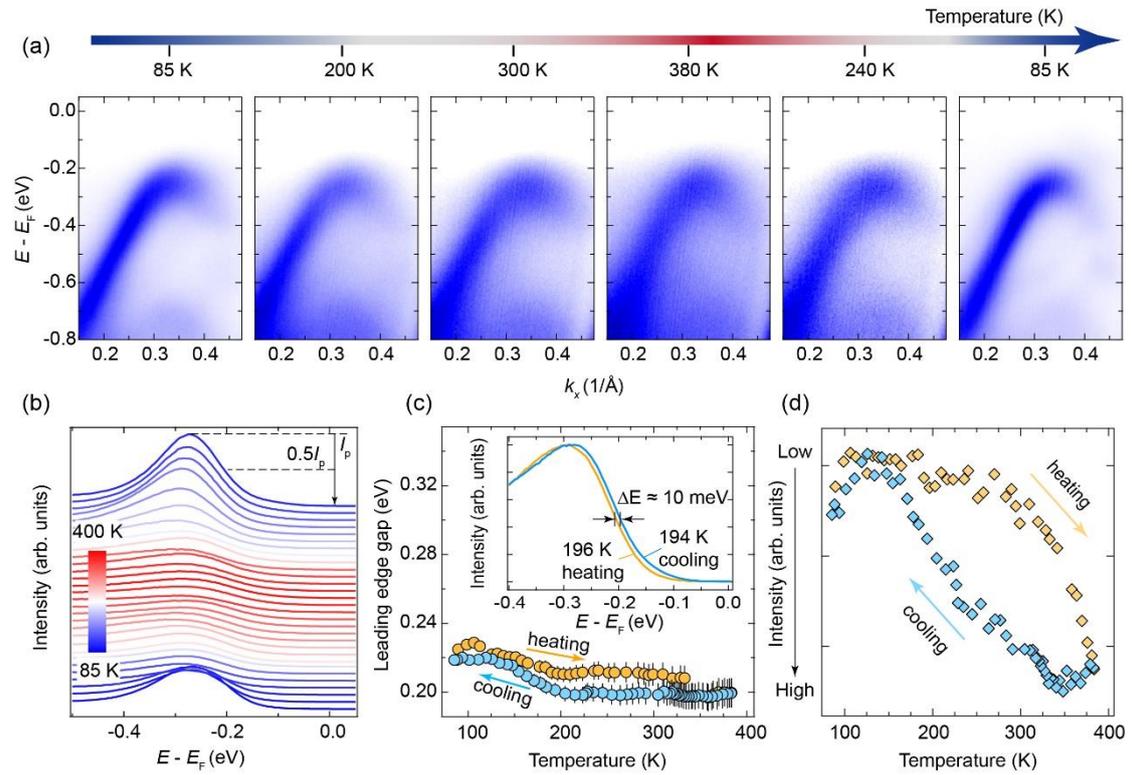

FIG.3. Temperature evolution of the band structure of EuTe$_4$. (a) Band dispersion along the $\bar{\Gamma}\bar{X}$ direction at selected temperatures. (b) Temperature evolution of the energy distribution curves (EDCs) integrated in a momentum window of about 0.08 Å$^{-1}$ around the momentum position of the band top. (c) Leading edge gap at $0.5I_p$ as a function of temperature, where $I_p$ is the maximum value of the EDC peak as marked in (b). The inset compares two typical EDCs at 196 K during heating up and 194 K during cooling down the sample. (d) ARPES intensity integrated in an energy window of 80 meV near $E_F$ as a function of temperature. Data were taken with 6.994 eV laser.

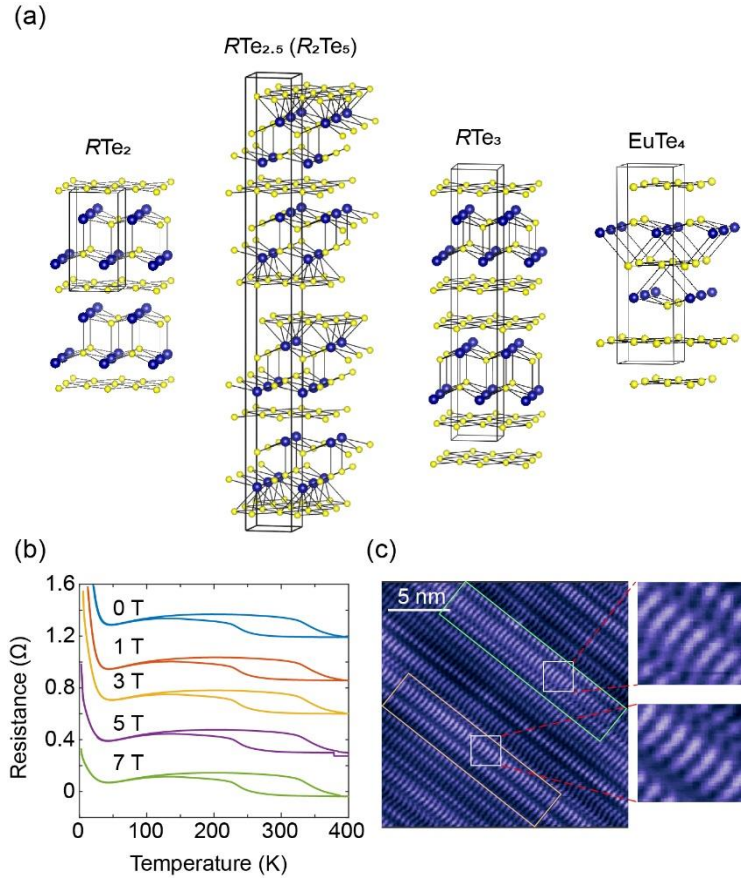

FIG.4. Possible mechanisms of the thermal hysteretic behavior in EuTe$_4$. (a) The crystal structure of $R$Te$_n$ compounds, where $R$ is a rare-earth element. (b) Temperature-dependent resistance of EuTe$_4$ under different magnetic fields. (c) STM surface topography of the top Te layer. The green and orange rectangles indicate the CDW domains with opposite trimerizing directions. The insets are the zoom-in plot of the data in the white squares.

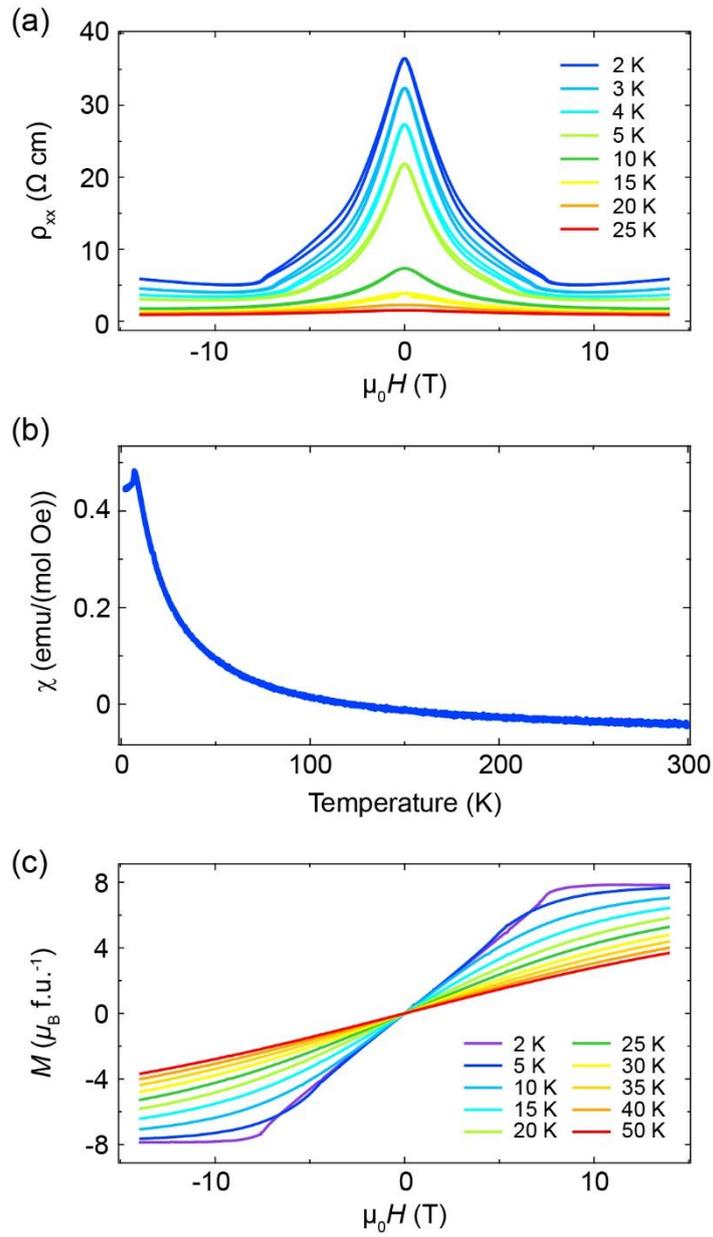

FIG.5. Magnetoresistance of EuTe$_4$. (a) Magnetoresistivity measured at different temperatures from 2 K to 25 K. (b) Magnetic susceptibility as a function of temperature. (c) Magnetization as a function of magnetic field measured at different temperatures.